\documentclass[reprint,aps,pra,showpacs,floatfix,twocolumn]{revtex4-1}
\usepackage{graphicx}
\usepackage{dsfont}
\usepackage{epstopdf}
\usepackage{amsmath}
\usepackage{amssymb}
\usepackage{color}

\newtheorem{theorem}{Theorem}
\newtheorem{lemma}{Lemma}

\newcommand{\bs}[1]{\boldsymbol{#1}}
\newcommand{\im}{\bs{\rm i}}

\newcommand{\e}{\bs{\rm e}}
\newcommand{\sgn}{\rm sgn}

\begin{document}

\title{Conditional limit measure of one-dimensional quantum walk with absorbing sink}

\author{Mohamed Sabri}
\affiliation{Graduate School of Information Sciences, Tohoku University, Aoba, Sendai 980-8579, Japan}

\author{Etsuo Segawa}
\affiliation{Graduate School of Information Sciences, Tohoku University, Aoba, Sendai 980-8579, Japan}

\author{Martin \v Stefa\v n\'ak}
\affiliation{Department of Physics, Faculty of Nuclear Sciences and Physical Engineering, Czech Technical University in Prague, B\v
rehov\'a 7, 115 19 Praha 1 - Star\'e M\v{e}sto, Czech Republic}

%\pacs{03.67.-a,05.40.Fb,02.30.Mv}

\date{\today}

\begin{abstract}
We consider a two-state quantum walk on a line where after the first step an absorbing sink is placed at the origin. The probability of finding the walker at position $j$, conditioned on that it has not returned to the origin, is investigated in the asymptotic limit. We prove a limit theorem for the conditional probability distribution and show that it is given by the Konno's density function modified by a pre-factor ensuring that the distribution vanishes at the origin. In addition, we discuss the relation to the problem of recurrence of a quantum walk and determine the P\'olya number. Our approach is based on path counting and stationary phase approximation.
\end{abstract}

\maketitle

\section{Introduction}

%%%%%%%%%%%%%%%%%%%%%%%%%%%%%%%%%%%%%%%%%%%%%%%%%%%%%%%%%%%%%%%%%%%%%%%%%%%%%%%%%%%%%%%%%%%%%%%%%%%%%%%%%%%

Quantum walks \cite{adz,fg,meyer} emerged during the 1990's as quantum mechanical extensions of classical random walks on a graph or a lattice, although similar ideas appeared already in 1960's in the works of Feynman and Hibbs on discretization of the Dirac equation \cite{feynman} and in the 1980's in the work of Gudder on quantum graphic dynamics \cite{gudder2,guder}. Since then they have attracted considerable attention due to their potential applications in quantum information \cite{aharonov}, e.g. in quantum search algorithms \cite{skw} or testing graph isomorphism \cite{iso:dou:08}.

Properties of quantum walks on infinite lattices are usually investigated in the asymptotic limit of large number of steps $n$. For homogeneous quantum walks on a line one can prove weak-limit theorems which show their ballistic spreading. The proofs are based either on path counting and combinatorial approach \cite{konno:limit:2002,konno:limit:2002b,konno:2005b}, or Fourier transformation \cite{grimmett}. The later method allows for straightforward extensions to quantum walks with larger internal degrees of freedom \cite{inui:grover1,miyazaki} (i.e. larger coin space) and higher-dimensional lattices \cite{watabe,hinajeros,MCH,MCHKB}. For quantum walks without translational invariance it is significantly more difficult to derive the explicit shape of the limit distribution, however, analytical treatment is still tractable in some cases. For example, the so-called CGMV method \cite{CGMV} which utilizes matrix-valued orthogonal Laurent polynomials \cite{BS} can be applied to quantum walks on a line \cite{KS} or a half-line \cite{cgmv:3} with position dependent coins.
Moreover, in terms of scattering theories, if the perturbation on the one-dimensional lattice is in the trace class, the convergence in law can be shown \cite{suzuki}.

In the present paper we consider a discrete-time quantum walk on a line where an absorbing sink is placed at the origin after the first step. The sink is modeled by a projection operator which sets the amplitude at the origin to zero. Hence, the overall evolution is not unitary. Quantum walks with sinks on various graphs were studied in the literature, focusing on the absorption probability \cite{ambainis:absorp,konno:absorp} and its time dependence \cite{bach:absorp,yamasaki:absorp,stef:perc}. We investigate the part of the wave-function which survives the absorption in the limit of infinite number of steps. It is shown that the conditional probability distribution to find the walker at position $j$, conditioned on the fact that the walker was not absorbed at the origin, converges weakly to a limit measure. Our proof is based on path counting approach \cite{FS} and stationary phase approximation \cite{Shiga}. 

The rest of the paper is organized as follows: we review the usual unitary two-state quantum walk on a line in Section~\ref{sec:prel} and set the notation. The main results of our paper are presented in Section~\ref{sec:sink}, where we state the limit theorem for the quantum walk with an absorbing sink at the origin. The properties of the conditional limit measure are illustrated on several examples. The detailed proof of the theorem is left for the Appendix~\ref{sec:app}. In Section~\ref{sec:rec} we discuss the relation of our results to the problem of recurrence and show that the conditional limit measure can be used to determine the P\'olya number of a quantum walk. We conclude in Section~\ref{sec:concl}.

%%%%%%%%%%%%%%%%%%%%%%%%%%%%%%%%%%%%%%%%%%%%%%%%%%%%%%%%%%%%%%%%%%%%%%%%%%%%%%%%%%%%%%

\section{Preliminary}
\label{sec:prel}

In this Section we set the notation and review the usual unitary quantum walk. We consider a two-state coined quantum walk on an integer lattice $\mathbb{Z}$. The Hilbert space of the walk is given by the tensor product of $\ell^2(\mathbb{Z})$ corresponding to the position of the walker on the lattice and $\mathbb{C}^2$ describing the internal coin state, i.e. 
%\ell^2(\mathbb{Z};\mathbb{C}^2)
\[ \mathcal{H} = \ell^2(\mathbb{Z})\otimes \mathbb{C}^2 =\{\psi: \mathbb{Z}\to \mathbb{C}^2 \;|\; ||\psi||_\mathcal{H}<\infty\}.  \] 
The scalar product in $\mathcal{H}$ is defined by 
	\[ \langle \varphi,\phi \rangle_{\mathcal{H}}=\sum_{j\in \mathbb{Z}} \langle \varphi(j),\phi(j) \rangle ,\]
where $\langle \cdot,\cdot \rangle$ denotes the standard scalar product in $\mathbb{C}^2$. The standard basis of $\mathbb{C}^2$ corresponds to the coin states $L$ and $R$, which determines the direction of the hopping of the walker to the left or to the right. Index $j$ labels the position of the walker on the lattice $\mathbb{Z}$. The two-component complex vector $\psi(j)\in\mathbb{C}^2$ consists of probability amplitudes of the walker being on the vertex $j$ with the coin states $L$ and $R$, i.e. $||\psi(j)||^2$ is the probability to find the walker at position $j$. The discrete-time evolution of the quantum walk is described by the unitary operator $U$. The operator $U$ can be decomposed into a product of the shift operator $S$ and the coin operator $C$. The coin operator $C$ does not change the position of the walker and it acts on the state $\psi$ as
$$
(C \psi)(j) = C_j \psi(j),
$$
where $C_j$ is a $2\times 2$ unitary matrix which can be position dependent, i.e. 
$$
C_j=\begin{pmatrix} a_j & b_j \\ c_j & d_j \end{pmatrix}.
$$
Throughout the paper we consider
\[ C_j=\begin{cases} C_+= \begin{pmatrix} a_+ & b_+ \\ c_+ & d_+ \end{pmatrix} & \text{: $j>0$,}\\ 
        \\
        C_0=\begin{pmatrix} a_0 & b_0 \\ c_0 & d_0 \end{pmatrix} & \text{: $j=0$,}\\ 
        \\
        C_-=\begin{pmatrix} a_- & b_- \\ c_- & d_- \end{pmatrix} & \text{: $j<0$,} \end{cases} \]
i.e. the coin operator $C$ can act in principle differently at the origin ($j=0$) and on the positive and negative semi-axis. We assume that the coin is mixing, i.e. $|a_{\pm,0}|\neq 0$ or 1.  The shift $S$ acts non-trivially on the position space $\ell^2(\mathbb{Z})$. Depending on the coin state ($L$ or $R$) it moves the walker to the left (i.e. from $j$ to $j-1$) or to the right (from $j$ to $j+1$). 
In total, the action of the operator $U$ on the state $\psi$ can be described by
\[ (U\psi)(j) = P_{j+1}\psi(j+1) + Q_{j-1}\psi(j-1),  \]
where 
	\[ P_j=\begin{pmatrix} a_j & b_j \\ 0 & 0 \end{pmatrix},\; Q_j=\begin{pmatrix} 0 & 0 \\ c_j & d_j \end{pmatrix},\  \mathrm{i.e.}\  P_j+Q_j = C_j . \]
The time evolution of the quantum walker according to the unitary operator $U$ is given by
$$
\psi_n = U \psi_{n-1} = U^n \psi_0.
$$
The subscript denotes the number of steps taken by the walker. For the initial state $\psi_0$ we consider the walker starting from the origin ($j=0$) with the coin state $\psi_c \in\mathbb{C}^2$, i.e. 
$$
\psi_0(j) = \begin{cases} \psi_c & \text{: $j=0$,}\\ & \\ 
0 & \text{: $j\neq 0$}. 
\end{cases}
$$
We write the initial coin state in the form
$$
\psi_c =  \left(\alpha,\ \beta\right)^T,
$$
and assume that it satisfies the normalization condition
$$
|\alpha|^2 + |\beta|^2 = 1.
$$
The subject of interest in quantum walks is the probability $\nu_n(j)$ to find the walker after $n$ steps at a position $j$ which is given by
$$
\nu_n(j) = ||\psi_n(j)||^2.
$$
The properties of the probability distribution $\nu_n(j)$ are well understood in the asymptotic limit $n\rightarrow\infty$. In particular, for the homogeneous case where $C_j = C_0$, we find the following limit theorem \cite{konno:limit:2002,konno:limit:2002b,grimmett,konno:2005b}:
\begin{theorem}
Let $X_n$ be the position of the walker after $n$ steps of the quantum walk, i.e. a random variable following the distribution $\nu_n$ which satisfies 
	\[ P(X_n\leq m) =\sum_{j\leq m} \nu_n(j).\]
Then we have 
	\[ \lim_{n\to\infty}P(X_n/n\leq y)=\int_{-\infty}^y \rho(x) dx  \]
for any $y\in \mathbb{R}$, where the limit density $\rho(x)$ is given by
$$
\rho(x) = (1-\lambda x) f_K\left(x;|a_0|\right).
$$
Here $f_K$ denotes the Konno's density function
\begin{equation}
\label{konno:df}
f_K(x;a)=\frac{\sqrt{1-a^2}}{\pi (1-x^2)\sqrt{a^2-x^2}},
\end{equation}
and the parameter $\lambda$ depends on the initial state $\psi_c$ according to
$$
\lambda = |\alpha|^2 - |\beta|^2 + \frac{a_0 \alpha \overline{b_0}\overline{\beta} + \overline{a_0}\overline{\alpha}b_0 \beta}{|a_0|^2} .
$$
\end{theorem}

We note that the limit density can be used to approximate the exact probability distribution according to the relation
$$
\nu_n(j) \approx \frac{2}{n}\rho\left(\frac{j}{n}\right),
$$
which is valid for $j$ and $n$ of the same parity. The factor of 2 is introduced artificially to offset the fact that the two-state walk is bipartite, i.e. $\nu_n(j) = 0$ for $j$ and $n$ of the opposite parity. The limit density captures the overall shape of the exact probability distribution and its main features, such as the positions of the peaks near $\pm|a_0| n$  which correspond to the divergencies of the limit density $\rho(x)$ for $x\rightarrow\pm |a_0|$.

\section{Quantum walk with an absorbing sink at the origin}
\label{sec:sink}

Let us now turn to the central issue of the paper, which is the quantum walk with an absorbing sink at the origin. If the walker returns to the origin it is annihilated by the sink and the walk ends. The sink sets the amplitude at the origin to zero while leaving the rest of the state untouched. Mathematically, sink is described by the projection operator $\Pi_0$ which acts on the state $\psi$ as
\[ (\Pi_0\psi)(j)=\begin{cases} \psi(j) & \text{: $j\neq 0$,}\\ & \\ 0 & \text{: $j=0$.} \end{cases} \]
Let us now consider the evolution of a quantum walk with the absorbing sink at the origin. The state of the walker after $n$ steps, provided that it has survived the action of the sink (i.e. it has not crossed the origin), is given by a non-normalized vector
\[ \psi^{(sur)}_n=(U^{(sur)})^n\psi_0. \]
Here we have defined 
$$ 
U^{(sur)}=(1-\Pi_0)U.
$$
The norm of the vector $\psi_n^{(sur)}$ is the survival probability up to $n$-th step
\begin{equation}
\label{p:surv}
P_n^{(sur)}=||\psi_n^{(sur)}||^2_{\mathcal{H}}=\sum_{j\in\mathbb{Z}}||\psi_n^{(sur)}(j)||^2 .
\end{equation}
Let us now define the conditional probability distribution, which is normalized by the survival probability, as
\[ \nu^{(cond)}_n(j)= \frac{||U\psi_{n-1}^{(sur)}(j)||^2}{P_{n-1}^{(sur)}}. \]
This is a quantum walk's analogue of the conditional probability on $\mathbb{Z}$ at time $n$ under the condition that the walker has never returned to the origin. 

The main result of our paper is the following limit theorem for the conditional probability distribution. 
\begin{theorem}
Let $X^{(cond)}_n$ be a position of the quantum walker after $n$ steps of the walk with an absorbing sink at the origin, i.e. a random variable following the distribution $\nu^{(cond)}_n$ and satisfying 
	\[ P(X^{(cond)}_n\leq m) =\sum_{j\leq m} \nu^{(cond)}_n(j).\]
Then we have 
	\[ \lim_{n\to\infty}P(X^{(cond)}_n/n\leq y)=\int_{-\infty}^y \rho^{(cond)}(x) dx  \]
for any $y\in \mathbb{R}$. The conditional limit density is given by 
$$
\rho^{(cond)}(x) = H(x)\rho^{(+)}(x) + H(-x)\rho^{(-)}(x),
$$
where $H(x)$ is the Heaviside step function and 
	\begin{align*}
        \rho^{(+)}(x) &= \frac{P_+}{N(|a_+|)} \ \frac{4x^2}{1+x}f_K(x;|a_+|), \\
        \rho^{(-)}(x) &=  \frac{P_-}{N(|a_-|)} \  \frac{4x^2}{1+|x|}f_K(x;|a_-|).
        \end{align*}
Here $f_K(x;a)$ is the Konno's density function and
\begin{eqnarray}
\nonumber P_+ & = & |\langle v_0^{(+)},\psi_c \rangle|^2,\quad \langle v_0^{(+)}|=(c_0,d_0) , \\
\nonumber P_- & = & |\langle v_0^{(-)},\psi_c \rangle|^2, \quad \langle v_0^{(-)}|=(a_0,b_0) .
\end{eqnarray}
By $N(a)$ we have denoted a normalization factor which reads
\[
N(a) = \frac{\pi a^2 - 2a\sqrt{1-a^2} + 2\left(1-2a^2\right) \arcsin(a)}{\pi  \left(1-a^2\right)}.
\]
\end{theorem}

We leave the detailed proof of the theorem for the appendix and turn to the discussion of the results. 

\begin{figure}
\includegraphics[width=0.48\textwidth]{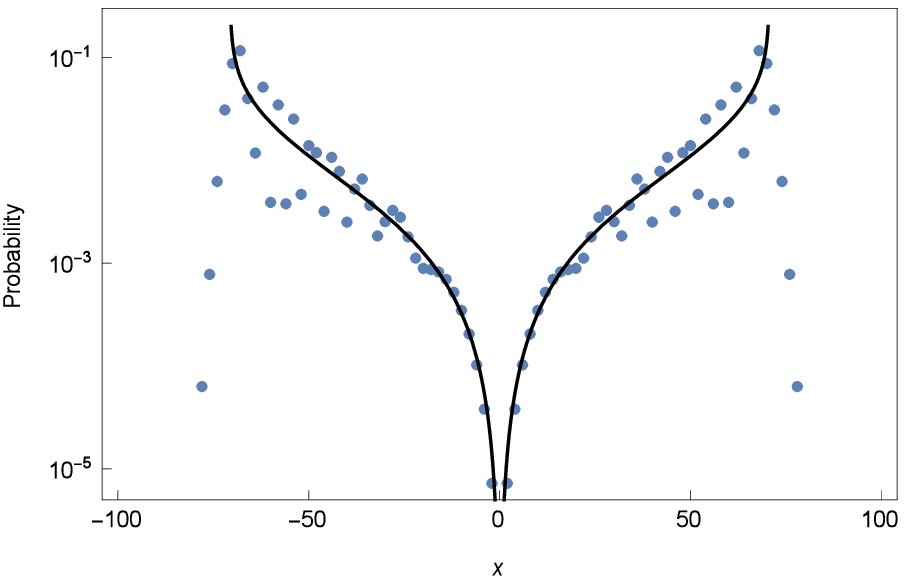} \hfill \includegraphics[width=0.48\textwidth]{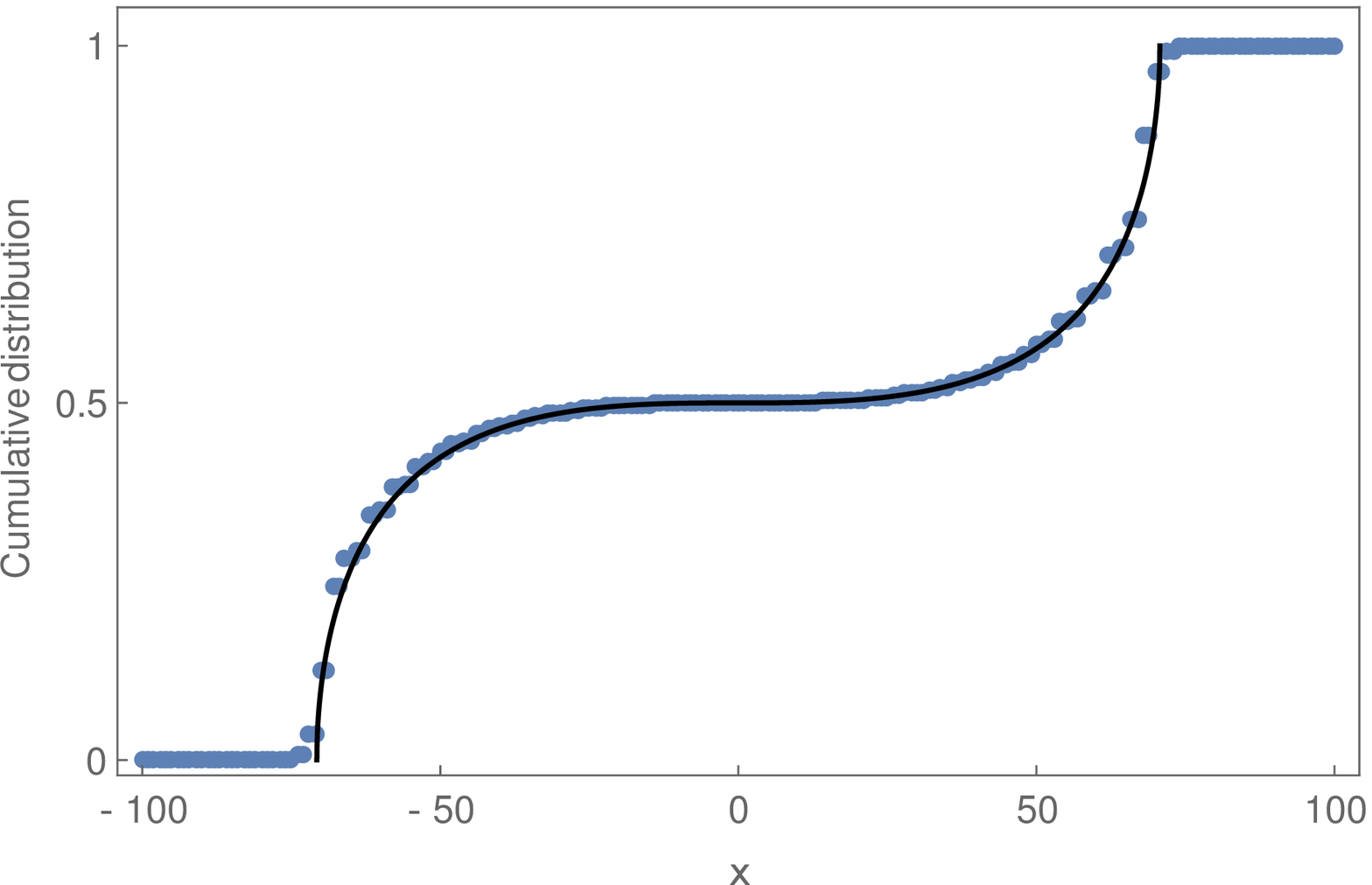}
\caption{In the upper plot we show the conditional probability distribution of the Hadamard walk after 100 steps with absorbing sink at the origin. Note that the plot is on the logarithmic scale. Blue dots are obtained from the numerical simulation, black curve is the approximation with the limit density of (\ref{cond:meas:had}). The initial coin state was chosen as $\psi_c = (1,0)^T$  which results in a symmetric conditional distribution. In the lower plot we present the cumulative distribution function where the rapid oscillations of the probability distribution are smeared out. The cumulative distribution function obtained from the limit density perfectly fits the exact numerical data.}
\label{fig:1}
\end{figure}

First, we see that the conditional limit measure is determined by the Konno's density function as for the walk without the absorbing sink, however, modulated by the factor $\frac{4x^2}{1+|x|}$. This contribution ensures that $\rho^{(cond)}(x)$ tends to zero for $x=0$ as expected due to the presence of the sink. We illustrate this feature in Figure~\ref{fig:1} where we consider the homogeneous case with the Hadamard coin, i.e.
$$
C_+ = C_- = C_0 = \frac{1}{\sqrt{2}} \begin{pmatrix} 1 & 1 \\ 1 & -1 \end{pmatrix}.
$$
As for the initial state we consider $\psi_c = (1,0)^T$, where the walker in the first step jumps to the left or to the right with equal probability, i.e. $P_+ = P_- = 1/2$. Hence, the conditional limit measure is symmetric and has the form
\begin{equation}
\label{cond:meas:had}
\rho^{(cond)}(x) = \frac{1}{2 - \frac{4}{\pi}} \frac{4x^2}{1+|x|} f_K\left(x;\frac{1}{\sqrt{2}}\right).
\end{equation}
We note that $P_\pm$ also gives the asymptotic conditional probability of finding the walker on the positive or negative half-line, i.e. 
\begin{eqnarray}
\nonumber P_+ & = & \int\limits_0^{|a_+|} \rho^{(cond)}(x) dx, \\
\nonumber P_- & = & \int\limits_{-|a_-|}^0 \rho^{(cond)}(x) dx .
\end{eqnarray}

\begin{figure}
\includegraphics[width=0.48\textwidth]{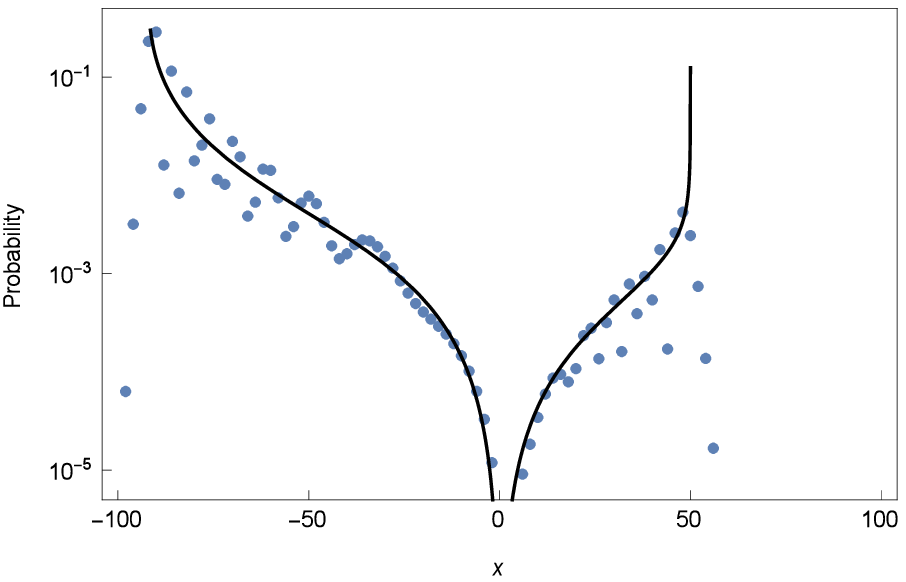} \hfill \includegraphics[width=0.48\textwidth]{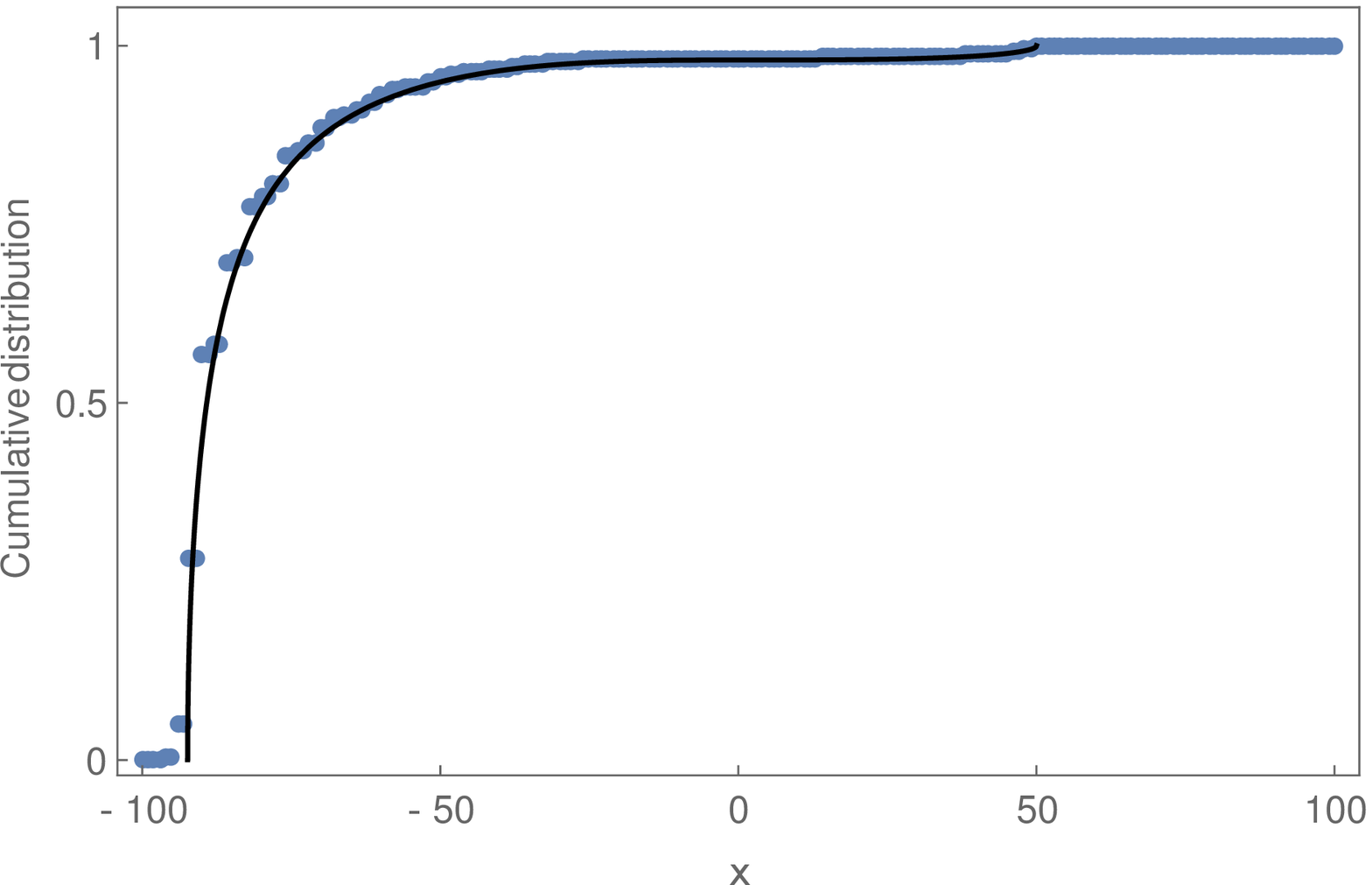}
\caption{Conditional probability distribution on the log-scale (up) and the cumulative distribution function (down) of the walk with absorbing sink at the origin. The initial coin state was $\psi_c = (0.6,0.8)^T$. The coins were chosen according to (\ref{coins:fig:2}).  Different coins for $j>0$ and $j<0$ result in asymmetric conditional distribution. In this case the walk spreads faster on the negative half-line than on the positive half-line. In addition, due to the choice of the initial coin state, the particle has a higher probability to be on the left side of the absorbing sink.}
\label{fig:2}
\end{figure}

As the next example we consider a case with different coins for the origin and the positive and negative half-line. The sink at the origin decouples the two half-lines. Hence, the evolution of the walker on the positive (negative) semi-axis is determined purely by $C_+$ ($C_-$). In Figure~\ref{fig:2} we have chosen the coins according to
\begin{eqnarray}
\label{coins:fig:2}
\nonumber C_+ & = & 
\frac{1}{2}\begin{pmatrix}
 1 & \sqrt{3} \\
 -\sqrt{3} & 1 \\
\end{pmatrix}, \\
 C_0 & = & \frac{1}{\sqrt{2}} \begin{pmatrix} 1 & 1 \\ 1 & -1 \end{pmatrix}, \\
\nonumber C_- & = & 
\begin{pmatrix}
 \cos \left(\frac{\pi }{8}\right) & \sin \left(\frac{\pi }{8}\right) \\
 -\sin \left(\frac{\pi }{8}\right) & \cos \left(\frac{\pi }{8}\right) \\
\end{pmatrix} .
\end{eqnarray}
Since $a_-=  \cos \left(\frac{\pi }{8}\right)> a_+ = \frac{1}{2}$, the walk spreads faster on the negative half-line than on the positive. Due to the choice of the initial state $\psi_c = (0.6,0.8)^T$ the conditional distribution is heavily biased towards left. Indeed, we find that $P_-=0.98$ and $P_+ = 0.02$, which is also clearly visible from the plot of the cumulative distribution function. 

\begin{figure}
	\includegraphics[width=0.48\textwidth]{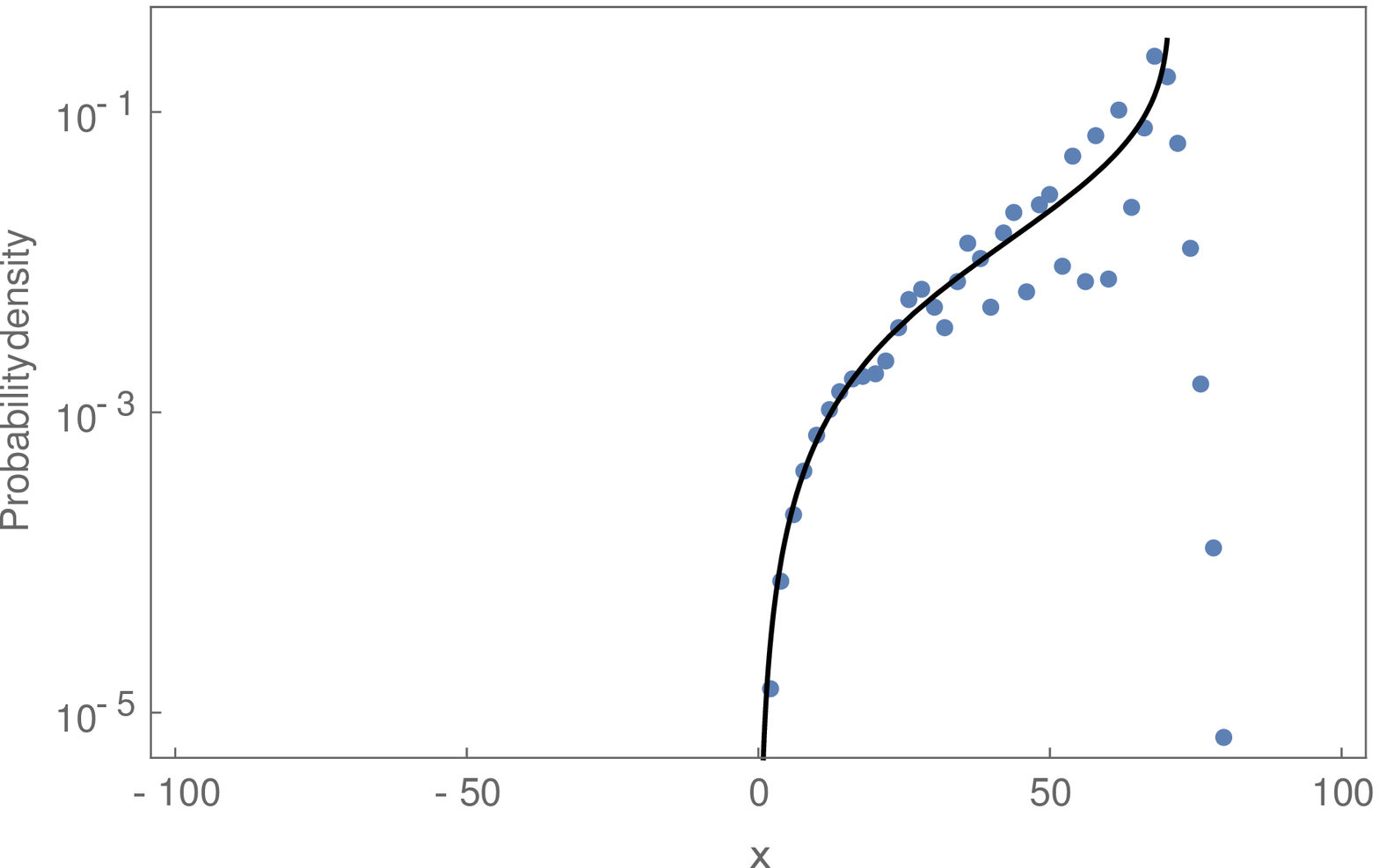} \hfill \includegraphics[width=0.48\textwidth]{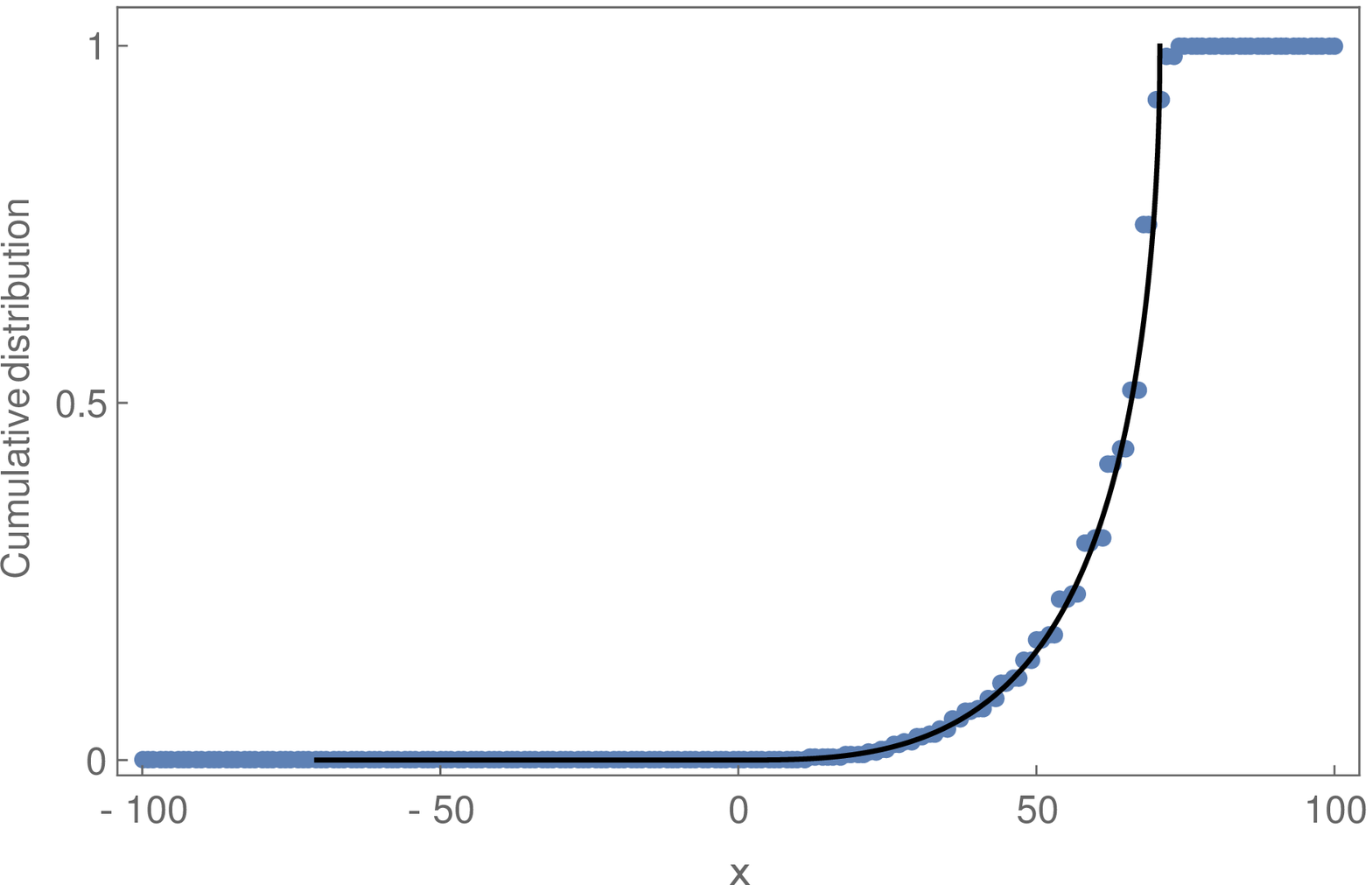}
\caption{Conditional probability distribution on the log-scale (up) and the cumulative distribution function (down) of the Hadamard walk with absorbing sink at the origin.The initial coin state was chosen as $\psi_c = \frac{1}{\sqrt{2}}(1,-1)^T$. In such a case the conditional distribution completely vanishes on the left side of the absorbing sink, since in the first step the walker jumps with probability 1 to the right and cannot cross the origin afterwards.}
\label{fig:3}
\end{figure}

Finally, we consider the case where the walker jumps in the first step with probability 1 to one side of the line, i.e. $P_+=1$ or $P_-=1$. This depends on the relation between the coin at the origin $C_0$ and the initial state $\psi_c$. From the unitarity of the matrix $C_0$ we find
\begin{eqnarray}
\nonumber P_- = 1 &\  \Longleftrightarrow\  & \psi_c = (\overline{a_0},\overline{b_0})^T, \\
\nonumber P_+ = 1 &\  \Longleftrightarrow \ & \psi_c = (\overline{c_0},\overline{d_0})^T.
\end{eqnarray}
Since the walker cannot cross the origin due to the presence of the sink the conditional probability distribution is non-zero only on the positive or negative half-line. We illustrate this feature in Figure~\ref{fig:3} for the case of the Hadamard walk. The initial state was chosen as $\psi_c = \frac{1}{\sqrt{2}}(1,-1)^T$ which results in $P_-= 0$ and $P_+ = 1$. Clearly, the walker can be found only on the positive half-line.

\section{Recurrence and P\'olya number}
\label{sec:rec}

Let us conclude by applying our results to the problem of recurrence \citep{polya}.  Here we are interested in the probability that the walker returns to the origin, which is called the P\'olya number. The walk is said to be recurrent if its P\'olya number is equal to one, and transient otherwise. For quantum walk there are two inequivalent definitions of the P\'olya number \cite{stef:rec,Grunbaum:recurrence} depending on the way the return to the origin is observed. In the observation scheme introduced in \cite{stef:rec} the measurement is performed only once after $n$ steps and the walk is restarted afterwards. The observation scheme in \cite{Grunbaum:recurrence} requires to monitor the origin continually after every step, and the walk ends if the walker is found. We note that in the classical case the two concepts are equivalent in the sense that recurrence under one definition implies recurrence under the other, and vice versa \citep{stef:rec2}. However, in the quantum case the two approaches lead to different results already for a quantum walk on a line, where in the restart scheme the walk is recurrent \citep{stef:rec2}, while in the continual scheme it is transient \citep{rec:kiss,ambainis:absorp}. Both observation schemes for recurrence were recently implemented in a photonic quantum walk experiment \citep{nitsche:exp} which has clearly shown the difference between the two schemes arising from the projective nature of quantum measurements.

Quantum walk with the sink at the origin investigated in this paper corresponds to the continual observation scheme for recurrence, where the return to the origin is marked by the absorption of the walker \citep{Grunbaum:recurrence}. In this scheme the P\'olya number of the walk is given by
$$
P = \sum\limits_{k=1}^\infty q(0,k),
$$
where $q(0,k)$ denotes the probability of the first return of the walker to the origin after $k$ steps. The first return probability is obtained from
$$
q(0,k) = ||(U\psi_{k-1}^{(sur)})(0)||^2.
$$
The sum 
$$
P_n^{(abs)} = \sum_{k=1}^n q(0,k),
$$
gives the probability that the walker was absorbed by the sink until the step $n$. Clearly, it is the complement of the survival probability (\ref{p:surv}), i.e.
$$
P_n^{(abs)} = 1 - P_n^{(sur)} .
$$
Since the P\'olya number of the quantum walk is given by the asymptotic value of the absorption probability, we find
$$
P = 1 - \lim\limits_{n\rightarrow\infty}  P_n^{(sur)}.
$$
As we discuss in the Appendix, the limiting value of the survival probability is given by
$$
\lim\limits_{n\rightarrow\infty}  P_n^{(sur)} = P_+ N(|a_+|) + P_-N(|a_-|).
$$
Hence, in the continual measurement scheme the P\'olya number for a two-state quantum walk on a line is given by
$$
P = 1 - P_+ N(|a_+|) - P_-N(|a_-|).
$$
We see that when the coin operator acts differently on the positive and negative half-line the P\'olya number depends on the initial coin state $\psi_c$ through the probabilities $P_\pm$. For $|a_+| = |a_-| = a$ the dependence on $\psi_c$ vanishes and we find that the P\'olya number is given by
$$
P = 2\frac{a \sqrt{1-a^2} + (1-2a^2)\arccos(a)}{\pi (1- a^2)},
$$
which coincides with the value obtained in \cite{Grunbaum:rec:subspace}. In the particular case of the Hadamard walk corresponding to $a=\frac{1}{\sqrt{2}}$ the P\'olya number reduces to $P = \frac{2}{\pi}$ in accordance with the findings of \cite{ambainis:absorp,rec:kiss}.

\section{Conclusions}
\label{sec:concl}

In this paper we have investigated the asymptotic properties of the quantum walk on a line with an absorbing sink at the origin. We have shown that the position of the quantum walker re-scaled by the number of steps $X_n^{(cond)}/n$ converges in the limit $n\to\infty$, similarly like for the unitary quantum walk. The explicit form of the conditional limit measure was derived. Finally, the relation to recurrence of a quantum walker was discussed. In particular, we have derived the P\'olya number in dependence on the initial state and the parameters of the quantum coin.

It would be interesting to investigate quantum walks with an absorbing sink on higher-dimensional lattices. For a one-dimensional lattice considered in the present paper the sink at the origin separates the dynamics into two independent quantum walks on a half-line. This is no longer true for higher-dimensional lattices. One can anticipate that the conditional limit density will exhibit more complicated patterns depending also on the initial state. However, this investigation is beyond the scope of the present paper. 

\begin{acknowledgments}
ES acknowledges financial supports from the Grant-in-Aid for Young Scientists (B) and of
Scientific Research (B) Japan Society for the Promotion of Science (Grant No.16K17637,
No.16K03939). 
M\v S is grateful for financial support from GA\v CR under Grant No. 17-00844S and from M\v SMT under Grant No. RVO 14000. 
\end{acknowledgments}

\appendix

\section{Proof of Theorem 2}
\label{sec:app}

Let $\Omega_n=\{1,-1\}^n$. For $\xi=(\xi_1,\dots,\xi_n)\in \Omega_n$, we define $\sigma_m=\xi_1+\cdots+\xi_m$ $(m\leq n)$. 
We define the weight of path: $w: \Omega_n\to M_2(\mathbb{C}_2)$ such that 
	\[ w(\xi)=P_{\xi_n,\sigma_{n-1}}\cdots P_{\xi_2,\sigma_1}P_{\xi_1,0} \]
for $\xi=(\xi_1,\dots,\xi_n)$, where 
	\[ P_{1,x}= Q_{{\sgn} (x)} \;\;\mathrm{and}\;\; P_{-1,x}=P_{{\sgn} (x)}. \]
We put $\Theta_j(n):=\{ \xi\in\Omega_n \;|\; \sigma_0=0,\;\sigma_1\neq 0,\dots,\sigma_{n-1} \neq 0,\;\sigma_n=j \}$ which is the set of 
all the possible paths from the origin to $j$ without the absorption. In the following we consider only the positive half-line $j\geq 1$, since the proof for the negative half-line is analogous. 
The weight of the paths avoiding to touch the absorbing sink is denoted by $\Xi_j^{(sur)}(n)=\sum_{\xi\in \Theta_j(n)} w(\xi)$. 
Remark that the non-normalized amplitude at position $j$ after $n$ steps can be obtained from
$$
\psi^{(sur)}_n(j)=\Xi_j^{(sur)}(n)\psi_c .
$$
Therefore, the weight $\Xi_j^{(sur)}(n)$ will play an important role. 
We introduce the generating function of $\Xi_j^{(sur)}(\cdot)$ with respect to time series $n$, i.e.
$$
\tilde{\Xi}_j^{(sur)}(z)= \sum_{n=0}^\infty \Xi_j^{(sur)}(n)z^{n},
$$ 
for a complex value $z$. We obtain the following lemma for the explicit expression of $\tilde{\Xi}_j^{(sur)}(z)$. 

\begin{widetext}

\begin{lemma}\label{Lem:gen}
Put $\e^{\im \delta}=\det(C_+)$, $\Sigma=\{\theta\in [0,2\pi) \;|\; |c_+|\geq |\sin\theta|\}$ and $\Sigma^c=[0,2\pi]\setminus \Sigma$. Let $\tilde{\Xi}_j^{(sur)}(z)$ be the above. Then we have 
	\begin{equation}\label{Eq:weight} 
        \tilde{\Xi}_j^{(sur)}(z)=\tilde{\lambda}^{j-1}(z)|\tilde{u}(z)\rangle\langle v_0^{(+)}| \;\;\; (j\geq 1), 
        \end{equation} 
where $|\tilde{u}(z)\rangle=(\tilde{\lambda}(z)\tilde{f}(z)\;,\;z)^T$ and $\langle v_0^{(+)}|=(c_0\;,\;d_0)$.
Here for $z=\e^{\im(\theta-\delta/2)}$, $\tilde{\lambda}(z)$ and $\tilde{f}(z)$ are given by
	\begin{align}
        \tilde{f}(\e^{\im (\theta-\delta/2)}) 
        	&= -e^{i\theta}\frac{|c_+|}{c_+}
                \begin{cases}
                \e^{\im \eta(\theta)} & \text{: $\theta\in \Sigma^c$,} \\
                \im \sgn(\sin\theta)\left( \left| \frac{\sin\theta}{c_+} \right|-\sqrt{\left| \frac{\sin\theta}{c_+} \right|^2-1} \right) & \text{: $\theta\in \Sigma$.}
                \end{cases} \label{f}\\ 
        \tilde{\lambda}(\e^{\im (\theta-\delta/2)})
        	&= \e^{\im\delta/2}\frac{|a_+|}{a_+}
                \begin{cases}
                \sgn(\cos\theta)\left( \left| \frac{\cos\theta}{a_+} \right|-\sqrt{\left| \frac{\cos\theta}{a_+} \right|^2-1} \right) & \text{: $\theta\in \Sigma^c$,} \\
                \e^{\im \kappa(\theta)}   & \text{: $\theta\in \Sigma$.}
                \end{cases} \label{lambda}
        \end{align}
where 
	\begin{align}
        \sin\eta(\theta) &= \frac{\sin\theta}{|c_+|},\;\;\sgn(\cos\eta(\theta))=\sgn(\cos\theta) \notag \\ 
        \cos\kappa(\theta) &= \frac{\cos\theta}{|a_+|},\;\;\sgn(\sin\kappa(\theta))=-\sgn(\sin\theta) \label{eq:theta}
        \end{align}
\end{lemma}
{\bf Proof of Lemma~\ref{Lem:gen}}: 
Replacing $\tilde{\Xi}_0(z)$ with $\tilde{\Xi}_0^{(sur)}(z)=I$ in Lemma~3.1 in \cite{KLS}, we obtain (\ref{Eq:weight}). 
The explicit expressions for $\tilde{f}$ and $\tilde{\lambda}$; (\ref{f}) and (\ref{lambda}), respectively, can be referred by (3.25) in \cite{KLS}.\;\;\;$\square$\\

We use the notation
$$
\tilde{\psi}_j^{(sur)}(z)=\tilde{\Xi}_j^{(sur)}(z)\psi_c.
$$
By the Cauchy's coefficient formula, 
the amplitude at time $n$ and position $j$ without the absorption $\psi_n^{(sur)}(j)$ is expressed by
	\begin{align*} 
        \psi_n^{(sur)}(j)
        &=\e^{\im n\delta /2}\int\limits_{0}^{2\pi} \tilde{\psi}_j^{(sur)}(\e^{\im (\theta-\delta/2)})\;\e^{-\im n\theta}\frac{d\theta}{2\pi} = \e^{\im n\delta /2}\left( I_j(n) + J_j(n) \right),
        \end{align*}
where we have introduced
$$ I_j(n) = \int\limits_{\Sigma}\tilde{\psi}_j^{(sur)}(\e^{\im (\theta-\delta/2)})\;\e^{-\im n\theta}\frac{d\theta}{2\pi} ,\quad    
J_j(n)  = \int\limits_{\Sigma^c}\tilde{\psi}_j^{(sur)}(\e^{\im (\theta-\delta/2)})\;\e^{-\im n\theta}\frac{d\theta}{2\pi} .
$$
Let us first consider $J_j(n)$. By Lemma~\ref{Lem:gen}, since $\theta\in \Sigma^c$, the absolute value of $\tilde{\lambda}$ is bounded as $|\tilde{\lambda}|<1$.
Therefore we have 
	\begin{align*}
        J_{j}(n) 
        = \langle v_0^{(+)},\psi_c\rangle 
        \int\limits_{\Sigma^c} \e^{-\im n\theta}\left\{\tilde{\lambda}(\e^{\im (\theta-\delta/2)})\right\}^{j-1} \;
        \begin{pmatrix} -\tilde{\lambda}(\e^{\im (\theta-\delta/2)})\; \e^{\im(\theta+\eta(\theta))}\;|c_+|/c_+ \\ \e^{\im(\theta-\delta/2)}\end{pmatrix} \frac{d\theta}{2\pi}\; \stackrel{(n\to \infty)}{\longrightarrow} \; 0,
        \end{align*} 
which is an exponentially decay to $0$. Therefore, $J_{j}(n) $ does not contribute to the limit density.
Let us turn to $I_j(n)$. In this case $\theta\in \Sigma$ and by Lemma~\ref{Lem:gen} $\tilde{\lambda}$ becomes a unit complex value. We find that for fixed $y\in\mathbb{R}_+$ the integral $I_{[ny]}(n)$ is expressed by 
	\begin{align*}
        I_{[ny]}(n) &= \e^{\im \sigma}\langle v_0^{(+)},\psi_c \rangle \int\limits_{\Sigma} 
        	\e^{\im \{ n(y\kappa(\theta)-\theta)+\beta_n(\theta)\}} |\tilde{u}(\e^{\im (\theta-\delta/2)})\rangle \frac{d\theta}{2\pi},
        \end{align*}
where $\beta_n(\theta)=([ny]-ny)\kappa(\theta)=O(1)$ and $\sigma=\delta/2-\arg(a)$. 
Here $|\tilde{u}\rangle$ is expressed by 
	\[ |\tilde{u}(\e^{\im (\theta-\delta/2)})\rangle= 
        	\begin{pmatrix} \e^{\im \omega(\theta)}\left( |\sin\theta/c_+|^2-\sqrt{|\sin\theta/c_+|^2-1}  \right) \\ \e^{\im(\theta-\delta/2)}\end{pmatrix}\] 
where 
	\[\e^{\im \omega(\theta)}=-\im \frac{|a_+||c_+|}{a_+c_+}\e^{\im \delta/2}\sgn(\sin\theta)\e^{\im(\kappa(\theta)-\theta)}.\] 	
We set $g(\theta):=y\kappa(\theta)-\theta$. 
The singular points of $g(\cdot)$ on $[0,2\pi)$ are described by $\{\theta_*^{(1)},\dots, \theta_*^{(4)}\}$. 
Now we can apply the following kind of extended stationary phase method. 
This lemma is obtained by following \cite{Shiga} even if 
we modify the function in the exponent by adding a function $\omega(x)$ with some conditions. 
We skip the proof in this paper. 
\begin{lemma}\label{lem:ESPM}(An extended stationary phase method)
Let $h:[x_*,b]\to\mathbb{R}$ be a $C^2$-function and 
$h'(x_*)=0$, $h''(x_*)\neq 0$ and $h'(x)\neq 0$. $(x_*<x\leq b)$. 
Let $\omega: [x_*,b]\times \mathbb{N}\to \mathbb{R}$ be a $C^1$-function with respect to $x\in [x_*,b]$, and satisfy 
the following condition
	\begin{align*} 
    \sup_{x,n}|w(x,n)|, \sup_{x,n}|w'(x,n)| &< c_1, \\
    |w(n,x)-w(n,x')| &< c_2|x-x'|,
    \end{align*}
where $c_1$ and $c_2$ are constant which are independent of $x$ and $n$. 
Then for $n\to\infty$ we find 
	\begin{align*}
    \int\limits_{x_*}^b \e^{\im (nh(x)+\omega(x,n))} dx
    = \frac{1}{\sqrt{n}}\;\sqrt{\frac{\pi}{2|h''(x_*)|}}\e^{\im \sgn(h''(x_*)\pi/4)}
    \e^{\im (nh(x_*)+\omega(x_*,n))}+ O(1/\sqrt{n}).
    \end{align*}
\end{lemma}
In our case Lemma~\ref{lem:ESPM} implies 
	\begin{align*}
    \e^{-\im \sigma}I_{[ny]}(n)/\langle  v_0^{(+)},\psi_c \rangle
    	&= \sum_{j=1}^4 
        \frac{1}{\sqrt{n}}\;\sqrt{\frac{2\pi}{|y\kappa''(\theta_*^{(j)})|}}
        \e^{\im \sgn(\kappa''(\theta_*^{(j)})\pi/4)}
        \e^{\im (n\kappa(\theta_*^{(j)})+\beta_n(\theta_*^{(j)}))}
        |\tilde{u}(\e^{\im (\theta_*-\delta/2)})\rangle \frac{1}{2\pi}+O(1/\sqrt{n}) .
    \end{align*}
Hence, for the square norm we find  
	\begin{align*}
        ||I_{[ny]}(n)||^2 
        = \frac{1}{n}\left(P_+ \; \sum_{j=1}^4 
        \frac{1}{|y\kappa''(\theta_*^{(j)})|}  ||\tilde{u}(\e^{\im(\theta_*^{(j)}-\delta/2)})||^2 \frac{1}{2\pi}
        +{\rm cross\;terms}\right)+O(1/n). 
        \end{align*}
Since $y\kappa'(\theta_*)-1=0$, using the formula of (\ref{eq:theta}) we obtain 
	\begin{align*}
        \frac{1}{\pi|y\kappa''(\theta_*^{(j)})|} &= y^2f_K(y;|a_+|), \\
        ||\tilde{u}(\e^{\im(\theta_*^{(j)}-\delta/2)})||^2 &= 1+|\tilde{f}(\e^{\im(\theta_*^{(j)}-\delta/2)})|^2 = \frac{2}{1+y}
        \end{align*} 
Since 
	\[ \int\limits_{0}^y ||I_{[nx]}(n)||^2dx = \frac{1}{n}\left(\sum_{j=1}^{[ny]} ||I_j(n)||^2+O(1)\right), \]
and the ``{\rm cross\;terms} " contains $``\e^{\im n(\theta_*^{(\ell)}-\theta_*^{(m)})+\beta_n(\theta_*^{(\ell)})-\beta_n(\theta_*^{(m)})}"$, 
we can use Lemma~\ref{lem:ESPM} again and find
	\begin{align*} 
    \sum_{j=1}^{[ny]}||I_{j}(n)||^2 
    &\stackrel{(n\to \infty)}{\longrightarrow} \int\limits_{0}^{y} P_+ \;\frac{4x^2}{1+x}f_K(x;|a_+|) dx.
    \end{align*}
This implies that 
$$
\lim\limits_{n\to\infty}\sum\limits_{j=1}^\infty ||\psi_n^{(sur)}(j)||^2  = P_+ N(|a_+|),
$$
where 
$$
N(|a_+|) = \int\limits_{0}^{|a_+|} \frac{4x^2}{1+x}f_K(x;|a_+|) dx.
$$
The conditional limit density for the positive half-line is then given by performing a proper normalization, i.e. 
$$
\rho^{(+)}(x) = \frac{P_+}{N(|a_+|)} \ \frac{4x^2}{1+x}f_K(x;|a_+|).
$$
For the negative half-line we proceed analogously and obtain the conditional limit density $\rho^{(-)}(x)$. We note that the limiting value of the survival probability (\ref{p:surv}) is given by
$$
\lim\limits_{n\rightarrow\infty}  P_n^{(sur)} = \lim\limits_{n\to\infty}\sum\limits_{j=1}^\infty ||\psi_n^{(sur)}(j)||^2 + \lim\limits_{n\to\infty}\sum\limits_{j=-1}^{-\infty} ||\psi_n^{(sur)}(j)||^2 =  P_+ N(|a_+|) + P_-N(|a_-|).
$$

\end{widetext}

\end{document}